# Technical Report

30 March 2019

*Passive Head-Mounted Display Music-Listening EEG dataset*

~

G. Cattan, P. L. C. Rodrigues, M. Congedo

GIPSA-lab, CNRS, University Grenoble-Alpes, Grenoble INP.
Address : GIPSA-lab, 11 rue des Mathématiques, Grenoble Campus BP46, F-38402, France

**Abstract** - We describe the experimental procedures for a dataset that we have made publicly available at https://doi.org/10.5281/zenodo.2617084 in mat (Mathworks, Natick, USA) and csv formats. This dataset contains electroencephalographic recordings of 12 subjects listening to music with and without a passive head-mounted display, that is, a head-mounted display which does not include any electronics at the exception of a smartphone. The electroencephalographic headset consisted of 16 electrodes. Data were recorded during a pilot experiment taking place in the GIPSA-lab, Grenoble, France, in 2017 (1). Python code for manipulating the data is available at https://github.com/plcrodrigues/py.PHMDML.EEG.2017-GIPSA. The ID of this dataset is *PHMDML.EEG.2017-GIPSA*.

**Résumé** - Dans ce document, nous décrivons une expérimentation dont les données ont été publiées sur https://doi.org/10.5281/zenodo.2617084 aux formats mat (Mathworks, Natick, USA) et csv. Cette base de données contient les enregistrements électroencéphalographiques de 12 sujets écoutant de la musique avec et sans casque de réalité virtuelle passif, c'est à dire, un casque n'incluant aucune électronique à l'exception d'un smartphone. Le casque d'électroencéphalographie comportait 16 électrodes. Les données ont été enregistrées durant une étude pilote réalisée au GIPSA-lab, Grenoble, France en 2017 (1). Un exemple d'application en Python pour manipuler les données est disponible sous le lien https://github.com/plcrodrigues/py.PHMDML.EEG.2017-GIPSA. L'identifiant de cette base de données est *PHMDML.EEG.2017-GIPSA*.

## Introduction

Virtual Reality (VR) is a promising field of application for Brain-Computer Interfaces (BCI) based on electroencephalography (EEG) as, in VR environments, BCI candidates to reduce the distance between a user and the virtual avatar. Passive head-Mounted Devices (PHMD) are VR devices consisting in a plastic or cardboard mask in which we insert a smartphone. They are cheap, affordable and thus may help to populate the VR+BCI technology for the general public, that is, they provide a potentially ubiquitous technology. However, to our knowledge, the impact of PHMD on the quality of EEG signal has never been assessed. We provide EEG recordings of subjects with and without a passive a head-mounted display while listening to classical music. A Fourier analysis (2) of these data has been published in (1). An example of application of this dataset is available online at https://github.com/plcrodrigues/py.PHMDML.EEG.2017-GIPSA.

## Participants

12 subjects participated to the experiment (3 females), with mean (sd) age 26.25 (2.63). Subjects were volunteers recruited at the University of Grenoble-Alpes. Before the experiment, the subjects were informed that they will be exposed to electromagnetic contamination due to the proximity of a smartphone put in front of their eyes. All participants provided written informed consent confirming the notification of the experimental process, the data management procedures and the right to withdraw from the experiment at any moment.

## Material

EEG signals were acquired using a research grade amplifier (g.USBamp, g.tec, Schiedlberg, Austria) and the EC20 cap equipped with 16 wet electrodes (EasyCap, Herrsching am Ammersee, Germany), placed according to the 10-10 international system (**Figure 1**). The locations of the electrodes were FP1, FP2, FC5, FC6, FZ, T7, CZ, T8, P7, P3, PZ, P4, P8, O1, Oz, and O2. The reference was placed on the right earlobe and the ground at the AFZ scalp location. The amplifier was linked by USB connection to the PC where the data were acquired by means of the software OpenVibe (3,4). We acquired the data with no digital filter and a sampling frequency of 512 samples per second.

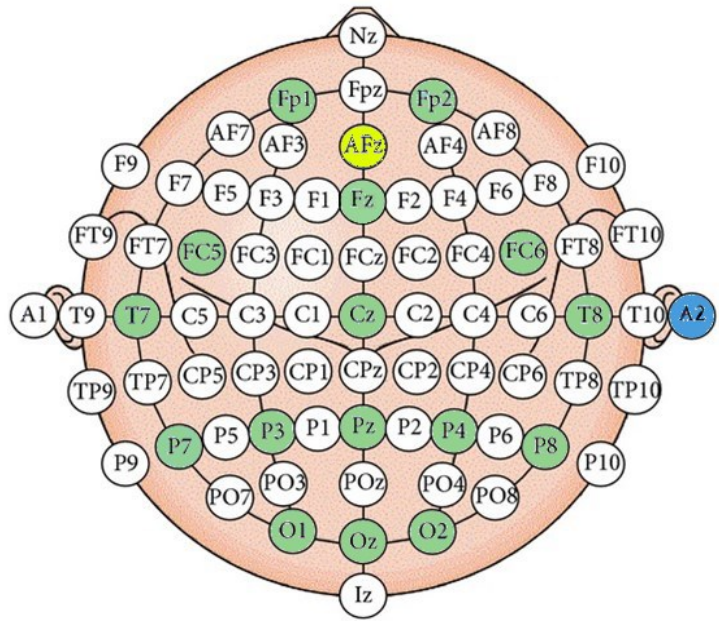

**Figure 1.** In green, the 16 electrodes placed according to the 10-10 international system. We used AFZ (in yellow) as ground and A2 (in blue) as a reference.

We used two identical smartphones in order to quickly switch between the two experimental conditions (with and without PHMD). In both conditions the subject wore a SamsungGear (Samsung, Seoul, South Korea) device (**Figure 2**). In one condition (without PHMD) the smartphone was switched-off, and in the other (with PHMD) it was switched-on. Anything else in the two conditions was identical. The smartphone used as VR devices was a Samsung S6 running under Android OS Nougat. The Specific Absorption Rate (SAR) of the smartphone was 0.382 Watt/Kg (Head) and 0.499 Watt/Kg (Body).

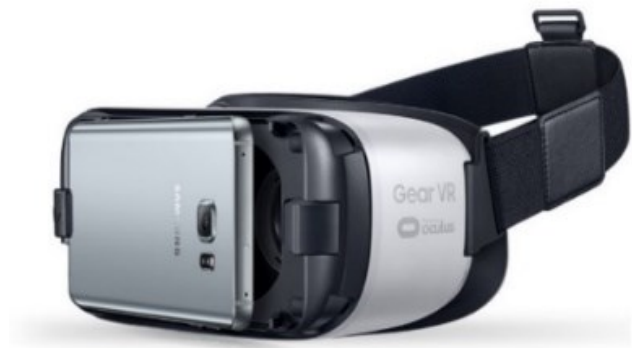

**Figure 2.** The SamsungGear (Samsung, Seoul, South Korea) consists in a plastic mask in which we insert a smartphone.

**Procedures**

Subjects were asked to sit on a desk in front of screen at a distance of about 50 cm. In order to mimic real-world usage we did not employ any instrumental noise-reduction device such as a Faraday cage. The EEG cap and the Samsung Gear were then placed on the subject. We continuously swapped the smartphones into the Samsung Gear (**Figure 3**). In both condition the screen of the smartphone was black and a purple marker was stuck on the left part of the screen in both devices. Having one marker might seems unnatural as one eye is looking at something that the other cannot. However, in a pilot study it was established that it was difficult for the subjects to reproduce stereoscopic vision with two markers because small differences in shape and position between the two markers were unavoidable. Additionally, there was a tiny white line on the middle of the switched-on smartphone to mark separation between left and right part of the screen on the running smartphone. This line was hidden by the Samsung Gear when the smartphone was put into it. The luminosity of the screen was comparable in the two conditions. Subjects were asked to focus on the marker and to listen to the music that was diffused during the experiment (Bach Invention from one to ten on harpsichord). The music was presented via the speakers of a personal computer. The marker and the music were introduced to homogenize the mental activity of the subjects during EEG recording. In addition, fixating the purple marker aimed at minimizing eye movement artifacts. The experiment comprised ten blocks. There were five blocks in the condition switched-on and five blocks in the condition switched-off (**Figure 3**). Each block consisted of one minute of EEG data recording with the eyes opened. Hence, a total of ten minutes were recorded for each subject. The sequence of the ten blocks were randomized prior to the experiment for each subject using a random number generator featuring no autocorrelation. This experimental design allows the use of exact randomization test for testing hypotheses (5). For ensuing analysis (1), we tagged the beginning of each block on the EEG using a trigger channel. Note that when the experimental condition was the same between two blocks, we removed and placed the same smartphone within the headset as if it were swapped.

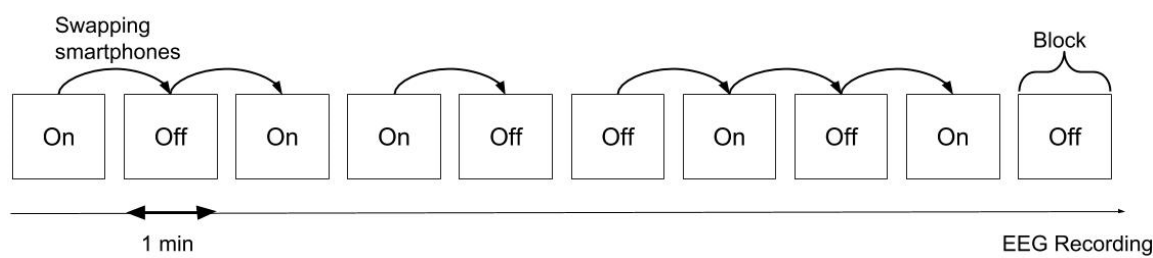

**Figure 3.** Example of a sequence comporting ten randomized blocs.

**Organization of the dataset**

For each subject we provide a single *mat* (and *csv*) file (Mathworks, Natick, USA) containing the complete recording of the session. The file is a 2D-matrix where the rows contain the observations at each time sample. Columns 2 to 17 contain the recordings on each of the 16 EEG electrodes. The first column of the matrix represents the timestamp of each observation and column 18 contains the triggers for the experimental condition. The rows in column 18 are filled with zeros, except at the timestamp corresponding to the beginning of a block, when the row gets a value of one (smartphone switched-off) or two (smartphone switched-on). The attribute names of the matrix are provided in the *Header.mat* (or *Header.csv*) file.

We supply an online and open-source example working with Python (6) and using the framework MNE (7,8). This example shows how to download the data using Python. The code shows also how to extract the blocks of given subjects from the dataset and plot the spectral characteristics of the signal.